\documentclass{ws-p9-75x6-50}

\newcommand{\mb}[1]{\mbox{\boldmath $#1$}}

\begin{document}

\title{Integrability conditions for vacuum spacetimes with a Killing
vector field}

\author{FRANCESC FAYOS}

\address{Departament de F\'{\i}sica Aplicada, UPC, E-08028 Barcelona,
Spain\\E-mail: labfm@ffn.ub.es}

\author{CARLOS F. SOPUERTA}

\address{Relativity and Cosmology Group, School of Computer Science and
Mathematics,\\
Mercantile House, Hampshire Terrace, PO1 2EG Portsmouth, England\\
E-mail: carlos.sopuerta@port.ac.uk}


\maketitle

\abstracts{We present a new approach to the study of vacuum spacetimes
with a Killing symmetry.  The central quantity in this approach is
the exterior derivative of the Killing vector field, which is a test
electromagnetic field.  Considering the algebraic structure of this
quantity we get a new view of the integrability conditions, which
provides a natural way of studying the connections between the algebraic
structure of the spacetime and properties of the Killing symmetry.}

Symmetries and in particular Killing symmetries have played an important
role in the search and study of exact solutions of Einstein's
equations.\cite{KSHM}   The main reason is that they lead to important
simplifications of the equations.  Another relevant way of simplifying
Einstein's equations is to assume a special algebraic structure of the
spacetime, which is determined by the Weyl tensor, and whose algebraic
classification is the Petrov classification.\cite{KSHM}
However, very little is known about the connections between Killing
symmetries and the algebraic structure of the spacetime.\cite{KSHM}
The aim of this communication is to present a new approach to the study
of non-null KVFs in vacuum space-times which is suitable to study such
connections.

The starting point is the fact, firstly noticed by
Papapetrou,\cite{PAPA} that, in vacuum, the exterior derivative
of a Killing vector field (KVF) $\mb{\xi}$ is a 2-form satisfying Maxwell's
equations in the absence of electromagnetic currents
\[ \mb{F}\equiv\mb{d\xi}\,,~~~~\nabla_{[a}F_{bc]}=0\,,~~
\nabla_{b}F^{ab}=0\,. \]
The main object that we use in our study is the algebraic structure of
$F_{ab}$.\cite{FASO}
There are only two different algebraic cases: (i) $F_{ab}$ is a
regular 2-form and hence it has two different principal null directions.
(ii) $F_{ab}$ is a singular 2-form having only one (multiple) principal
null direction.  The case $F_{ab}=0$ corresponds either to the Minkowski
spacetime or {\em pp waves} (when the KVF is a null vector field).

The first application is to combine the algebraic structures of $F_{ab}$ and
of the Weyl tensor $C_{abcd}$ in order to classify space-times having at
least one KVF, or more precisely, the pairs $\{(V_4,\mb{g}),\mb{\xi}\}$,
which takes into account the fact that there are space-times with more than
one KVF. Then, we can classify these pairs according to the following
properties: The algebraic type of $F_{ab}$ [(i) or (ii) above]; the Petrov
type of the space-time (I, II, III, D, N, or O); the
degree of alignment of the principal directions of $F_{ab}$ with those of
the Weyl tensor $C_{abcd}\,.$ Finally, we can refine this classification
by adding differential invariant properties of the principal null directions
of $F_{ab}$ and $C_{abcd}\,.$

The second application is to set up a new formalism for vacuum space-times
with an isometry.  The idea is to make an extension of the Newman-Penrose
(NP) formalism, a formalism suitable to control the algebraic
structure of the spacetime. This extension has two ingredients: (A) To
write all the equations in a NP basis in which $F_{ab}$ takes its canonical
form [(i) or (ii) above]. (B) To add the following variables and equations:
The components of
the KVF $\mb{\xi}$ and $F_{ab}$ in such a basis.  The equations for these
quantities are: the definition of $F_{ab}$ for the components of $\mb{\xi}$
and the Maxwell equations for the components of $F_{ab}\,.$
Then, the alignment of a principal direction of $F_{ab}$ with one of
$C_{abcd}$ can be study in a natural way within this formalism.  For instance,
setting $\Psi_0=0$ we impose a principal direction of $F_{ab}$ to be
aligned with one principal direction of the space-time.

In this formalism the integrability conditions for the components
of the KVF determine completely the Weyl tensor as an algebraic combination
of spin coefficients and quantities constructed from the KVF.\cite{FSTS}
As a consequence, we do not need to solve the second Bianchi identities,
which are the equations for $C_{abcd}\,.$ Instead, we only
have to substitute $C_{abcd}$ in them, obtaining a set of consistency
relations.  Moreover, it turns out that the whole set of consistency relations
and remaining integrability conditions constitute additional equations for
the spin coefficients, which complement the Newman-Penrose equations.
Therefore, we have to study their compatibility.  The case in which $F_{ab}$
is singular has been completely examined\,\cite{FSTS} and all the spacetimes
and the KVFs determined:  They correspond to two particular classes of
{\em pp waves} (Petrov type N solutions) and the Minkowski space-time.
Currently we are studying the regular case.\cite{FSCO}
As an example, in the case of Petrov type III vacuum space-times we
have arrived to the conclusion\,\cite{FSTS} that the alignment of the
multiple principal direction of $C_{abcd}$ with any of the two principal
directions of $F_{ab}$ is forbidden.   In contrast to this situation,
there are other vacuum space-times in which we can find alignments.
An interesting example is the case of the Kerr metric in which the two
multiple principal directions of the space-time (it is Petrov type D) are
aligned with those of the Papapetrou field.\cite{FASO}

\section*{Acknowledgments}
F.F. acknowledges financial support from the D.G.R. of the
Generalitat de Catalunya (grant 1998GSR00015), and the Spanish Ministry of
Education (contract PB96-0384). C.F.S. is supported by the European
Commission (contract HPMF-CT-1999-00149).

\end{document}